\documentclass[fleqn]{article}
\usepackage{graphicx}
\usepackage[figuresright]{rotating}
\usepackage{axodraw}
\usepackage[T1]{fontenc}
\begin{document} 
{\centering
{\Large {\bf
Characterization of the JUDIDT Readout Electronics for
Neutron Detection}

\vspace{1.0cm}  
{\large R.\ Fabbri$^{a, b}$, U.\ Clemens$^a$, 
R.\ Engels$^a$, G.\ Kemmerling$^a$, and \mbox{S.\ van Waasen$^a$}}\\

\vspace{0.3cm}      
{\large 
$^a$ Forschungszentrum J\"{u}lich (FZJ), J\"{u}lich \\
$^b$ Corresponding Author: r.fabbri@fz-juelich.de} \\
\vspace{1.5cm}
      \hspace{3.7cm} 
\centering{\today}
}
}

\vspace{2.5cm}
\begin{abstract}
  The Group for the development of neutron and gamma detectors in the 
  Central Institute of Engineering, Electronics and Analytics 
  (ZEA-2) at Forschungszentrum J\"{u}lich (FZJ) has developed, 
  in collaboration with
  European institutes, an Anger Camera prototype for improving the 
  impact point reconstruction of neutron tracks. The detector is a 
  chamber filled with $^3He$+$CF_4$ gas for neutron capture and 
  subsequent production of a tritium and a proton. 
  The energy deposition by the ions gives rise to drifting electrons 
  with an avalanche amplification as they approach a  
  micro-strip anode structure. The scintillating light, generated 
  during the electron drift and avalanche stage, is collected by four 
  vacuum photomultipliers. The position reconstruction is performed via 
  software algorithms. The JUDIDT readout electronics was modified at 
  ZEA-2 to cope with the data acquisition requirements of the prototype. 
  The results of the commissioning of the electronics are here presented 
  and commented.
\end{abstract}

{\large \vspace{-18cm} \hspace{-4cm} 
  Forschungszentrum J\"{u}lich\\

  \hspace{-4.cm}
  Internal Report No. FZJ\_2013\_02194
}

\newpage  

  \tableofcontents
  \vspace{-20cm} \hspace{6cm} 
\newpage  
\section{Introduction}
%
At current and future neutron facilities, like the research reactor 
FRM-II~\cite{FRM-II}, or the European Spallation Source ESS~\cite{ESS}, 
Small Angle Neutron Scattering (SANS) will obtain a major improvement 
by the much higher neutron fluxes than previously delivered. This leads 
to an intensity gain of at least one order of magnitude for SANS 
instruments, allowing a more detailed study of complex structures and 
fast processes. To cope with these expected high intensities
the new detector readout system system JUDIDT~\cite{JUDIDT} was developed 
for the SANS-diffractometer KWS1 at the previously active research reactor 
FRJ-2 of the Forschungszentrum J\'{u}lich, Germany. The system was then 
modified for a later use at the KWS2 experiment at FRM-II. 

Within the FP7/NMI3~\cite{FP7} European-founded project it was proposed to 
use the JUDIDT readout electronics for the data acquisition of a novel 
Anger Camera prototype~\cite{ANGER}, being able to cope with counting rates 
of several hundred kilohertz. Some modifications were implemented to 
properly handle the signal generated by the scintillating light in the 
$^3He$ volume gas. 
The proposed prototype underwent a full series of measurements at several 
neutron test-beam facilities in Europe, showing the capability to cope 
with the high demanding performance~\cite{ANGER}.

In this paper the results of the characterization
of the stand-alone electronics are presented and commented. In addition, the
data acquisition with the entire system detector plus the JUDIDT readout
system will be described and the corresponding results discussed.

\section{JUDIDT Electronics: Description and Properties}
%
\label{sec:JUDIDT}
To cope with the high rates provided by the modern neutron beam facilities,
the ZEA-2 Institute at FZJ, has developed the readout system JUDIDT 
(J\"{u}licher Digitales Detektor Auslesssystem). The design of the 
electronics comprises a fast signal and data processing by usage of 
modern technologies like field-programmable gate arrays (FPGAs). 

\begin{figure*}[t!]
    \hspace{-1.3cm}
      \includegraphics[height=8cm,width=14.cm]{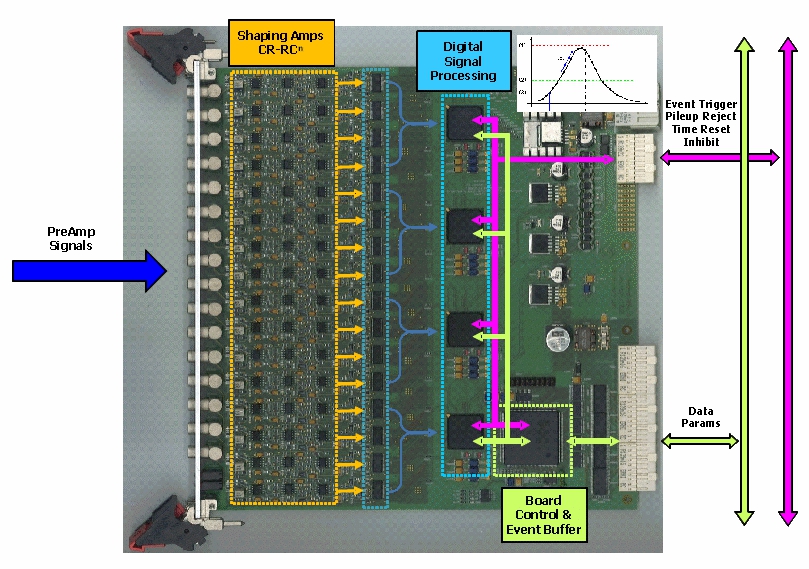}
  \caption{Components of the JUDIDT module: with different color 
           codes are highlighted different stages of the signal 
           processing: the sixteen $50$ Ohm LEMO connectors for input 
           signal, the amplifying and shaping stage, the ADCs, and 
           the FPGAs.}
  \label{fig:JUDIDT_schematics}
\end{figure*}

A detailed description of the electronics can be found in~\cite{JUDIDT}. 
Here we address only its main features: the front-head is equipped with
LEMO $50$ Ohm connectors to handle up to $16$ pre-amplified signals.
According to the selected capacitor on the integrator stage the 
amplifier, different gain can be obtained. 
The signal leaves the main amplification and shaping stage 
with a decay time typically of $400$ ns, and is then feeded to 
its dedicated 12-bit ADC in the board, Fig.~\ref{fig:JUDIDT_schematics}. 
Each FPGA handles four input channels and stores in a circular buffer
the values provided by the individual ADCs, which sample their corresponding
signal at a rate of $40$ MHz (every $25$ ns). 
As soon as a maximum in one channel is found, then the amplitude is 
evaluated also for the other channels in a selectable time window 
enclosing the measured peaking time. In the main FPGA register several 
cuts can be set to select events in specific amplitude and time windows. 
The setting of the FPGA is performed via API functions provided
by the driver developers.

The $12$ bit ADC (AD9236) covers an amplitude range from zero up to 
two Volts with an ADC resolution of $0.488$ mV per ADC unit. The 
data in the upper half of the ADC dynamic range ($11$ bits)
are then shipped to an external computer through a $1$ GByte optical 
link using only $8$ bits, thus deteriorating the signal resolution 
down to $3.9$ mV. Considering that the main amplification stage is 
configured to provide a gain factor of ten results an overall resolution 
measured by the analyzer of $0.39$ mV per ADC unit.

%

\section{Test-Bench Description}
\begin{figure*}[t!]
      \includegraphics[height=6cm,width=12.cm]{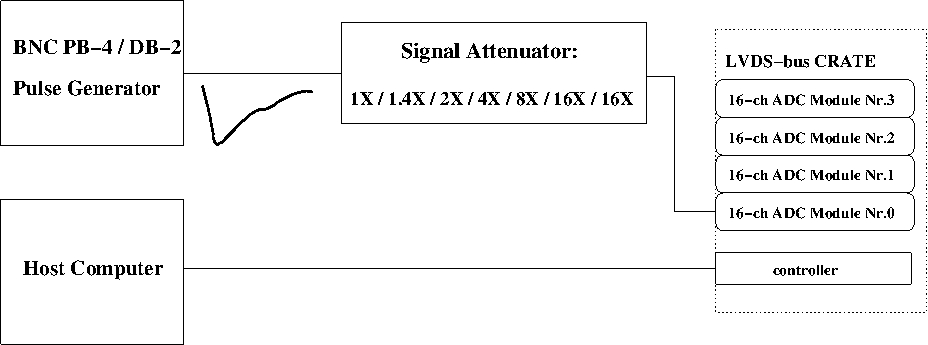}
%
  \caption{Typical test-bench setup used for the characterization of the 
           JUDIDT electronics at FZJ.} 
  \label{fig:TestBench}
\end{figure*}

%
The characterization measurements of the JUDIDT electronics were 
performed at the laboratories in ZEA-2,
The typical setup of the test-bench is shown in 
Fig.~\ref{fig:TestBench}. 
An analogue voltage generator (BNC by Berkeley Nucleonics Corp.) 
provides a tailed pulse with negative polarity for different selectable 
values of pulse rate and with variable amplitude. Additionally, a 
device by Tennelec was used to provide selectable attenuator factors 
to the voltage tailed signal. 

The attenuated signal is then injected in one of the $16$ input 
channels of the investigated board, and the processed output signal 
(e.g., the measured peaking amplitude) is saved in a binary file on 
hard disk, using the dedicated data acquisition software developed for 
the laboratory~\cite{DAQ}. The data acquisition is
driven from a Windows XP machine connected to the LVDS-bus based crate 
(develped at ZEA-2) via a SIS1100 Gbit
optical link interface on both sides of the link. The DAQ can be easily ported
to a LINUX system, being the software cross-platform build. 
The electronics performance has been also investigated using it as readout
system for the Anger Camera Prototype~\cite{ANGER}, using 
$Cf^{256}$ neutron sources in the ZEA-2 laboratory. 
%

%

\section{Electrical Noise Investigation}
%
The readout system has a very low intrinsic noise. Due 
to this optimal feature the pedestal distribution in each 
channel appears like a delta function at one or seldomly 
two ADC units. In this scenario it is difficult to reliably 
estimate the system noise as the root mean squared of 
the pedestal distribution.

A possible solution is to couple one input channel to the 
negative unipolar output of an amplifier. Here, we have used 
the amplifier ORTEC $571$. Different gain factors were applied
in order to make the pedestal distribution broader, thus covering
as many ADC channels as possible. It is clear that while 
coupling an amplifier to the readout electronics the measured
noise is somehow the convolution of the noise arising from 
both systems, and can likely provide only an upper limit 
estimation.

\begin{figure*}[t!]
      \includegraphics[height=7cm,width=12.cm]{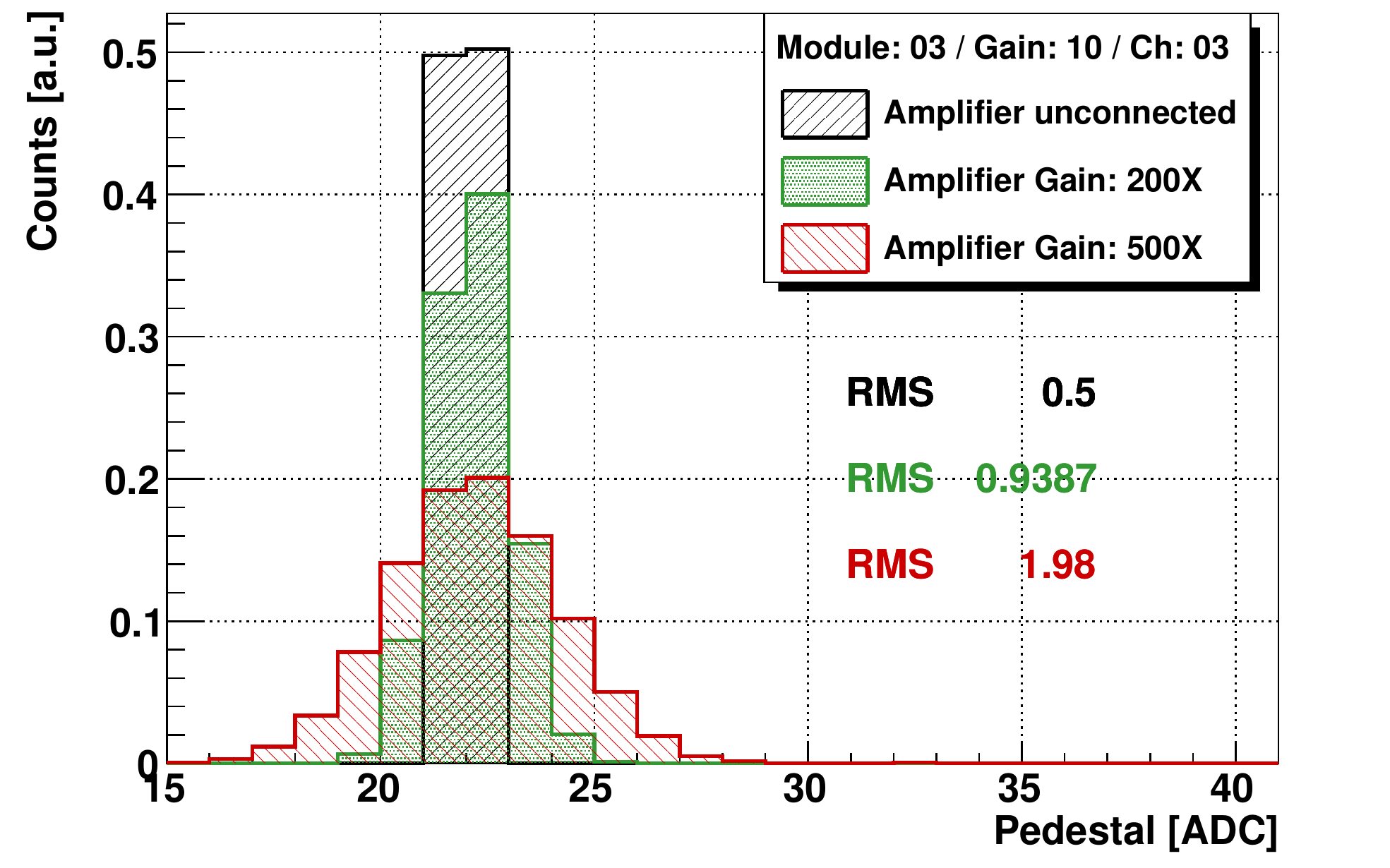}
  \caption{Pedestal distribution obtained with different configurations 
           for the ORTEC amplifier as described in the text.} 
  \label{fig:Noise}
\end{figure*}

An example of our procedure is presented in Fig.~\ref{fig:Noise},
where the pedestal of the same channel is shown in different 
configurations for the amplifier. 
At the largest gain used, the distribution is well Gaussian 
and covers many ADC channels, with a spread of 1.98 ADC units,
resulting in a negligible noise level.
%
%

\section{Pedestal Uniformity and Adjustment}
\begin{figure}[t!]
  \includegraphics[height=7.5cm,width=13.5cm]{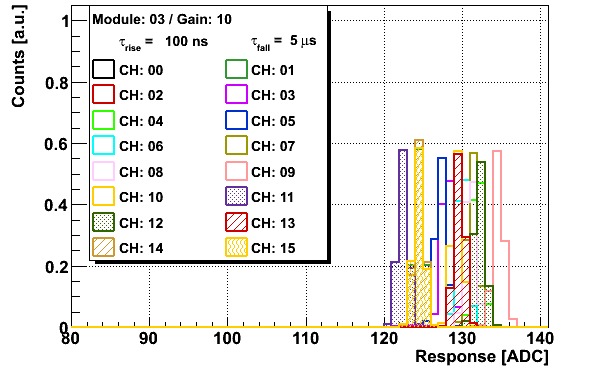}
  \vspace{-0.5cm}
  \caption{A tailed signal with unchanged amplitude is measured separately 
          in all the $16$ input channels of the module. As explained in
          the text a non uniform response can be observed and corrected.}
  \label{fig:SignalVsCh}
\end{figure}

%
The peaking amplitude of the input signal can be measured
differently by the
$16$ independent channels of a board. This effect is due by the different
intrinsic pedestal offset of each channel, and by a possible gain mismatch of
the individual amplification channels. An example of this bias is shown in
Fig.~\ref{fig:SignalVsCh}, resulting in an effective covered dynamic 
range not equivalent for all the $16$ ADC channels.

\begin{figure}[t!]
 \includegraphics[height=7.5cm,width=12.cm]{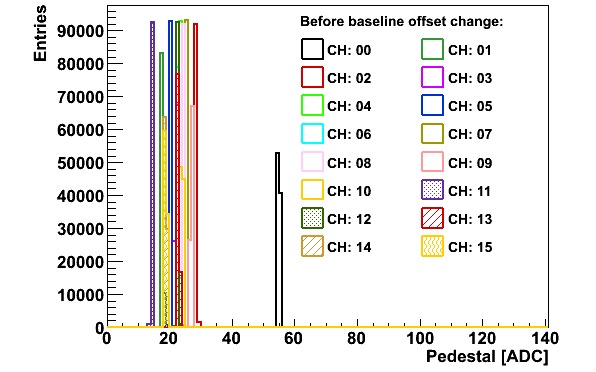}
 \includegraphics[height=7.5cm,width=12.cm]{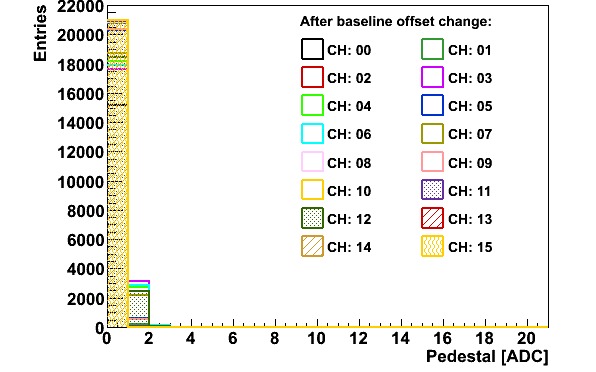}
  \caption{Pedestal distribution for all the $16$ channels
           without (top panel) and with baseline correction 
           (bottom panel) as described in the text.}
  \label{fig:PedestalSpread}
\end{figure}

This feature can be cured by changing the baseline offset which can be modified
by the data acquisition software tuning the DAC at the end of the amplifying
stage, Fig.~\ref{fig:JUDIDT_schematics}. The effect of the baseline correction
is presented in Fig.~\ref{fig:PedestalSpread}. The
gain mismatch in the amplifying stage and/or in the voltage setting of the
photomultiplier can be recovered instead by gain-matching procedures, as the
technique described in Sec.~\ref{sec:GAIN_MATCHING}.

It is clear that by following this procedure the entire ADC range can be
covered by all input channels, avoiding artificial bias in the measurements.
%

\section{Cross-Talk between Input Channels}
\label{sec:CROSS_TALK}
In a multi-channel readout system it is important to verify that 
the signal cross-talk between the individual channels remains at 
a negligible level. Here the investigation of the cross-talk is 
done performing a wider scan of injected charge in a single channel. 
While increasing the injected charge in several steps up to the maximum 
ADC value, the signal, that is the pedestal, of the remaining channels 
is measured and monitored for variations.  

\begin{figure}[t!]
 \includegraphics[height=7.5cm,width=13.cm]{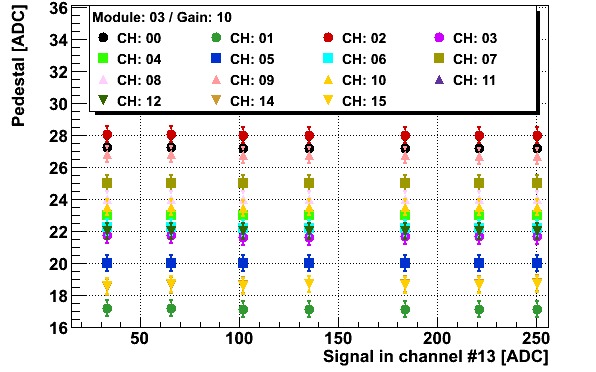}
  \caption{While increasing the amplitude of the signal in channel $13$
           the pedestal is measured in the remaining channels of the module.}
  \label{fig:CrossTalk}
\end{figure}

An example of the performed measurements is presented in 
Fig.~\ref{fig:CrossTalk}. The measured
pedestal is presented for all channels while in the channel $13$ the charge was
injected with different amplitude values. In the neighboring lines the
measured signal appears to remain well within one ADC unit. 

\begin{figure}[t!]
 \hspace{-2.3cm}
 \includegraphics[height=7.5cm,width=9.cm]{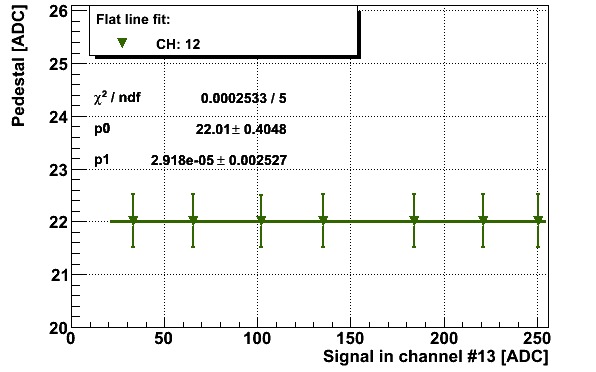}
 \hspace{-0.8cm}
 \includegraphics[height=7.5cm,width=9.cm]{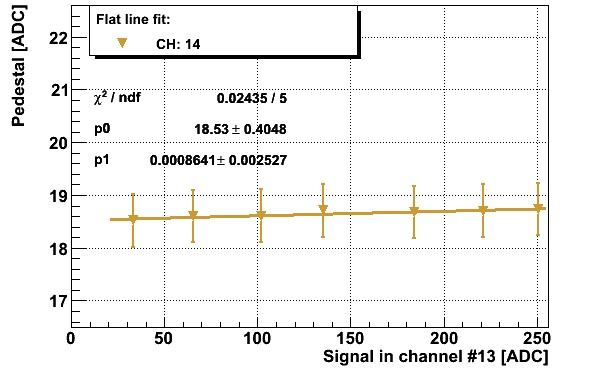}
  \caption{Example of estimation of the cross-talk, performed as 
           described in the text.}
  \label{fig:CrossTalkFit}
\end{figure}

An attempt to
estimate a possible variation is presented in Fig.~\ref{fig:CrossTalkFit}, 
where for each channel a
two-parameter linear fit is performed in the analyzed input signal range,
confirming a negligible charge leakage between the channels. A small trend is
possibly visible in channel $14$, which is processed by the same FPGA of the
channel $13$, where the signal was injected. Conservately, for each data point
the uncertainty was taken as being half of the ADC bin size. This assumption
is motivated by the narrow pedestal distribution, which is typically well
within one ADC unit. 
%

\section{Dynamic Range and Linearity of the Readout Electronics}
%
During data taking the dynamic range of the ADC should linearly cover the
entire possible energy range deposited by the impinging neutrons. In the 
Anger Camera the scintillating light is initially generated during the 
drift of the generated electron cloud in the gas volume and then in the 
amplification stage of the micro-strip. At the exit window of the camera 
the light is collected by a set of vacuum photomultipliers (four PMTs in 
the prototype used at FZJ), and converted in a current signal. 
The amplitude of the signal depends eventually
on several factors as the tension applied to the micros-trip, to the drift
field, and to the photomultipliers, and the amplifying stage of the JUDIDT
electronics, which is set to ten for this type of experiment. The electronics gain
can be eventually changed in order to cover the dynamic range allowed by the
investigated physics process. This can be achieved by using a proper
capacitor in the main amplification stage.

Special care should be taken to the preamplifier configuration to fit the
characteristic of both the incoming signal from the PMTs and of the JUDIDT
amplifying stage. An input signal of $120$ mV can already saturate the
amplifier, when the device is configured for a gain of factor ten, 
(as verified directly on the board by a scope).

The linearity of the system response was tested with the laboratory data
acquisition system and using a negative tailed pulse at different amplitude
values. The minimum injected voltage was the reference point of the
measurement, and was extracted at the oscilloscope
after maximally attenuating the signal with the Tennelec device. This device
offers several combinations of attenuating values by properly activating the 
several switches available in the device. By de-selecting those
switches a larger amplitude could be injected in the readout electronics
without modifying the setting of the pulse generator. 

\begin{figure}[t!]
  \includegraphics[height=8.cm, width=13.0cm]{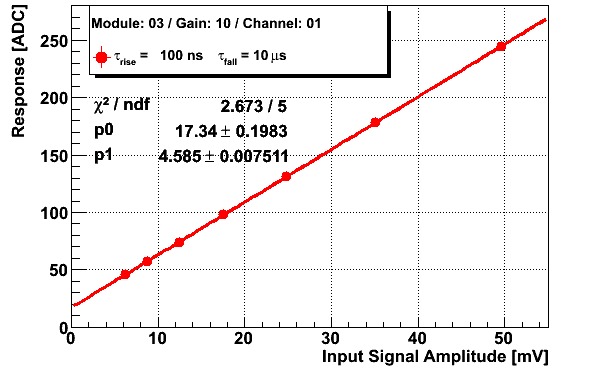} 
  \includegraphics[height=8.cm, width=13.0cm]{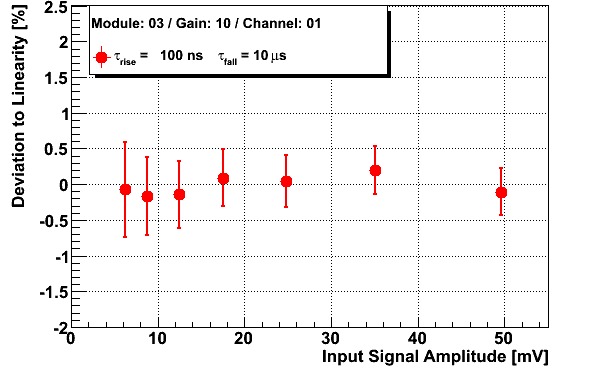} 
  \caption{  Top Panel: A linear fit is performed to the experimental data to
    verify the linearity of the electronics response. Bottom Panel: residuals
    to linearity calculated as described in the text.}
  \label{fig:LINEARITY}
\end{figure}

An example of this measurement is presented in the upper panel of
Fig.~\ref{fig:LINEARITY} for tailed pulses with $10$ $\mu$s decaying time. 

A linear fit to the experimental data is performed; for each measurement setup
a Gaussian distribution is generated, and the uncertainty on the mean value is
calculated, following~\cite{PDB}, as $RMS/\sqrt(N-1)$. The residuals to the 
linearity are calculated dividing the deviation of the measured experimental 
points from the linear fit results over the fit values. The results are 
presented as percent in the  bottom panel of Fig.~\ref{fig:LINEARITY}, 
showing a linearity 
well stable within $0.5\%$ within the statistical uncertainty of the 
measurement.

It should be noted that by using an analogue pulse generator the minimum 
injected voltage (our reference) could not be measured with enough precision. 
This could possibly affects the extraction of the fit slope, but not any 
conclusion about the linearity of the system, being the larger input voltage
points determined by rescaling the measured reference point using the 
switches of the Tennelec device.

The system appears therefore reasonably linear, while keeping the input 
signal below the saturation region of the main amplifying stage, above 
mentioned. The gain of factor $10$ of the electronics was verified 
directly on the board before the ADC stage via a high-impedance scope.

\section{Gain of the Readout Electronics and ADC Calibration}
%
Although great care is taken in the design a construction of each 
independent channel to have the same signal processing,
eventually slight differences in the gain are observed, as shown in 
Tab.~\ref{tab:Gain}. Here, a signal of $40$ mV was injected separately 
in each channel of two modules with nominal gain $10$, and the 
corresponding amplified and shaped signal was measured at the end of the 
amplifying stage before the ADC module, using a high-impedance (1 MOhm)
probe. In the module Nr.03 (Nr.04) a drop in the gain up to $7.7\%$
($4.9\%$) can be observed 
with respect to the maximum value, and a maximum deviation with 
respect to the nominal value up to $10\%$ ($2\%$); thus verifying 
that indeed a non-homogeneous gain among channels and boards as 
well is present.
\begin{table}[t!]
  \begin{center}
\begin{tabular}{||l|c|c|c|c|c|c|c|c||}
\hline 
\hline
\multicolumn{9}{|c|}{ \textbf{Module 03} }\\
\hline
\hline
\bf{Channel:}     & 00  & 01  & 02   & 03  & 04   & 05   & 06  & 07  \\
\bf{Signal [mV]:} & 390 & 395 & 400  & 390 & 405  & 400  & 395 & 395 \\
\bf{Gain:}        & 9.7 & 9.9 & 10.0 & 9.7 & 10.1 & 10.0 & 9.9 & 9.9 \\
\hline 
\hline 
\bf{Channel:}     & 08   & 09   & 10  & 11   & 12   & 13   & 14  & 15  \\
\bf{Signal [mV]:} & 400  & 400  & 395 & 410  & 405  & 400  & 390 & 390 \\
\bf{Gain:}        & 10.0 & 10.0 & 9.9 & 10.2 & 10.1 & 10.0 & 9.7 & 9.7 \\
\hline
\hline
\multicolumn{9}{|c|}{ \textbf{Module 04} }\\
\hline
\hline
\bf{Channel:}     & 00  & 01  & 02  & 03  & 04  & 05  & 06  & 07  \\
\bf{Signal [mV]:} & 390 & 360 & 360 & 370 & 370 & 370 & 370 & 370 \\
\bf{Gain:}        & 9.7 & 9.0 & 9.0 & 9.2 & 9.2 & 9.2 & 9.2 & 9.2 \\
\hline 
\hline 
\bf{Channel:}     & 08  & 09  & 10  & 11  & 12  & 13  & 14  & 15  \\
\bf{Signal [mV]:} & 360 & 370 & 370 & 360 & 370 & 365 & 375 & 360 \\
\bf{Gain:}        & 9.0 & 9.2 & 9.2 & 9.0 & 9.2 & 9.1 & 9.4 & 9.0 \\
\hline
\end{tabular}
\caption{Different gain values are measured in the channels of both 
         boards Nr.03 and Nr.04, configured to provide a nominal 
         gain of factor $10$.}
\label{tab:Gain}
  \end{center}
\end{table}

This feature can be easily cured by properly adjusting the high voltage  
of the involved photo-multipliers, after a gain matching procedure, 
as for example, described in Sec.~\ref{sec:GAIN_MATCHING}.

\begin{figure}[t!]
  \includegraphics[height=8.cm, width=13.0cm]{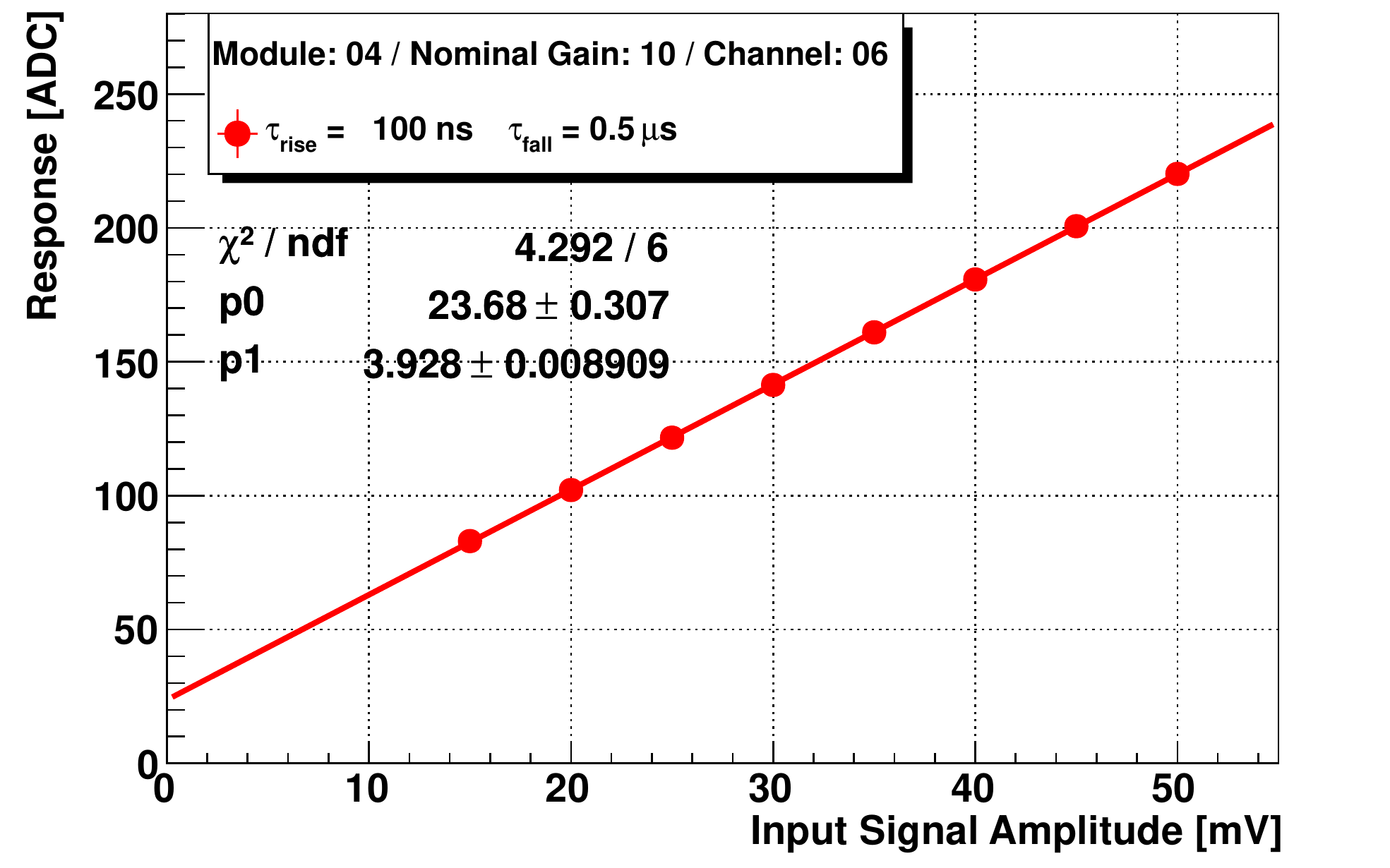} \\
  \vspace{-0.6cm}
  \caption{The system resolution is extracted for a specific channel
           following the procedure described in the text.}
  \label{fig:ADC_CALIB}
\end{figure}

From simple considerations in Sec.~\ref{sec:JUDIDT} an estimation for the 
signal resolution was obtained as $0.39$ mV per measured ADC for a board
with ideally a gain of $10$. As we have proved, this is not the case,
because the gain can vary largely between channels and boards.
Therefore it is interesting to measure the signal resolution 
directly in the test bench facility. Using the digital pulse generator 
BNC/PB-5 the resolution was directly measured injecting pulses 
with increasing and well known amplitude. An example of this analysis 
is presented in Fig.~\ref{fig:ADC_CALIB}.

Performing a linear fit to the experimental data, from the inverse of the
extracted slope the overall system ADC resolution of $1./3.928 = 0.25$ 
\mbox{mV / ADC} can be obtained for 
the analyzed channel. This measurement confirms again the high linearity
of the readout system.

\section{Rate Efficiency}
\begin{figure}[t!]
 \includegraphics[height=8.5cm,width=13.cm]{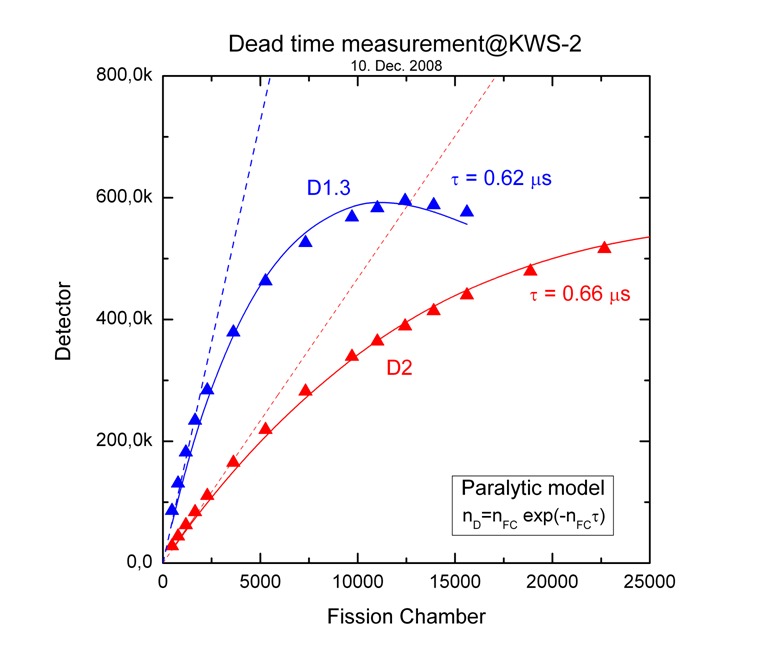}
  \vspace{-1.0cm}
  \caption{Measurement of the detector dead time performed at the experiment
           KWS-2 at FRM-II.}
  \label{fig:RATES_KWS}
\end{figure}

%
The DAQ system is forseen to cope with high rate fluxes of some hundreds
of kilohertz, and has been therefore tested to fullfil this requirement. An
interesting measurement was in the past done in the experimental hall at the
KWS-2 detector in FRM-II. Taking as reference the counting of a fission
chamber, the detector KWS-2, using the JUDIDT electronics, was located at two
different locations with respect to the flux exit window. The result of this
measurement is presented in Fig.~\ref{fig:RATES_KWS}. 

\begin{figure}[t!]
 \includegraphics[height=7.cm, width=12.cm]{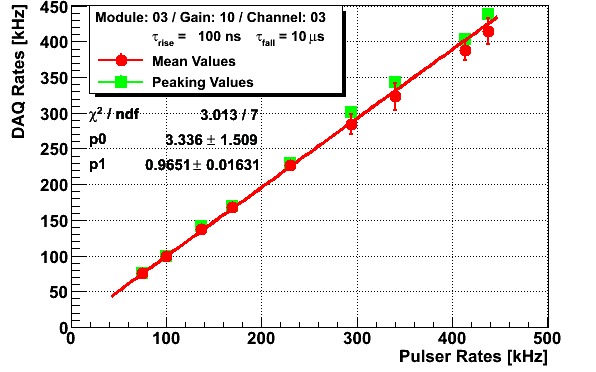}
 \vspace{-0.4cm}
  \caption{Using the DAQ developed for the laboratory, the rate efficiency 
           is measured using a tailed pulse generator operated at different 
           frequency values.}
  \label{fig:RATES_LAB}
\end{figure}

At FZJ a similar measurement has been performed using a tail pulse generator
operated at different frequency values, and storing the data with the
timestamp granularity of one millisecond, Fig.~\ref{fig:RATES_LAB}. 

It is clear that following this
technique no statistical Poisson fluctuation in the flux can be observed, as
instead is expected in a reactor beam. The efficiency is calculated for each
run, and its mean is evaluated considering several runs. The displayed
uncertainty is obtained considering the maximum deviation between all
considered runs. The points at large rate values show larger uncertainty,
which can be temptatively understood as given by the not sufficiently small
timestamp granularity. Superimposed to the points is shown a linear fit, whose
slope appears to be driven downwards by the high rate points, possibly due to
the fact that the multi-purpose DAQ developed in the laboratory is not yet
optimized for this specific measurement. The peaking rate values, 
superimposed in the plots, are typically larger than the mean efficiency 
values, thus supporting this scenario. 

\begin{figure}[t!]
  \includegraphics[height=7.cm, width=12.cm]{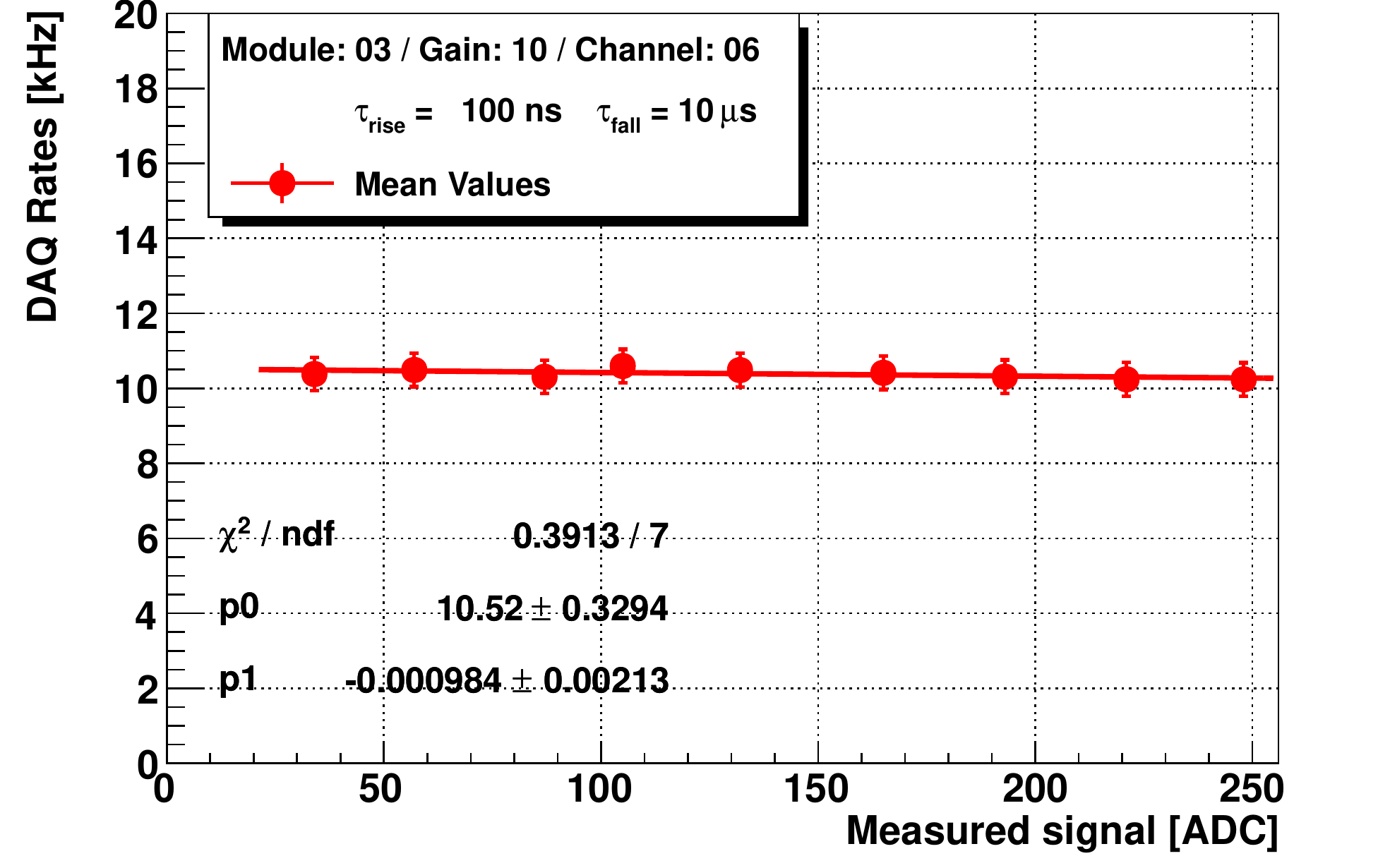}\\
  \vspace{-0.6cm}
  \caption{A signal with varying amplitude is injected at fixed pulse rates.
           The linear fit, as described in the text, do not 
           show any dependence (at this used signal rate) 
           of the data acquisition
           on the signal rates.}
  \label{fig:RATE_UNIFORMITY}
\end{figure}

It is additional interesting to investigate also how homogeneous 
is the data accumulation rate of the electronics with respect to 
the amplitude of the 
incoming signal. This feature was studied by injecting a tailed 
signal with varying amplitude and fixed pulse rate while measuring 
the rates with the laboratory DAQ system. No precise amplitude 
value could be determined directly by the front panel of the 
analogue BNC generator; nevertheless this uncertainty should not
have an effect on this analysis. An example of this 
measurement is presented in Fig.~\ref{fig:RATE_UNIFORMITY}
For each injected signal configuration the DAQ rate is 
calculated consecutively each $2000$ accumulated events, 
and the ADC mean of the several measurements is shown for each 
signal amplitude. The presented uncertainty is only systematics, 
and represents conservately half of the maximal deviation of the 
observed rate values (deviations due to the load of 
computing time of the laboratory DAQ when performing 
internal routines such as, among the others, listening at 
clients or dumping the running configurations).    
A two-parameter linear fit is performed to the experimental 
data showing no anomalous trend in the acquisition rate.
%

\section{Gain Matching of the Photomultipliers}
%
\label{sec:GAIN_MATCHING}
In order to correct for the above mentioned gain mismatch
among the different readout channels in the system, comprising
the photomultipliers and the amplifying electronics, a 
blue light LED system has been implemented to equalize the response of 
all PMTs. The results presented in this paper were obtained 
using the Hamamatsu vacuum photmultipliers R580 which have 
a nominal photo-cathode diameter of $38$ mm. 

The LED is centrally located in a hole in the front-head  
structure, where the PMT photocathodes (symmetrically located 
with respect to the LED) watch at the exit window of 
the Anger Camera (where the scintillating light is 
generated). 
Blue light is emitted by energizing the LED with a positive 
voltage pulse with a width and amplitude of approximately $300$ ns 
and $3$ V. The emitted light does not reach directly the photocathodes, 
but can reach them only upon isotropic backward reflexion 
on the chamber window, thus allowing for a reasonably
uniform illumination of the photomultiplier photocathodes.

\begin{figure}[t!]
 \hspace{-2.5cm}
 \includegraphics[height=6.5cm, width=8.5cm]{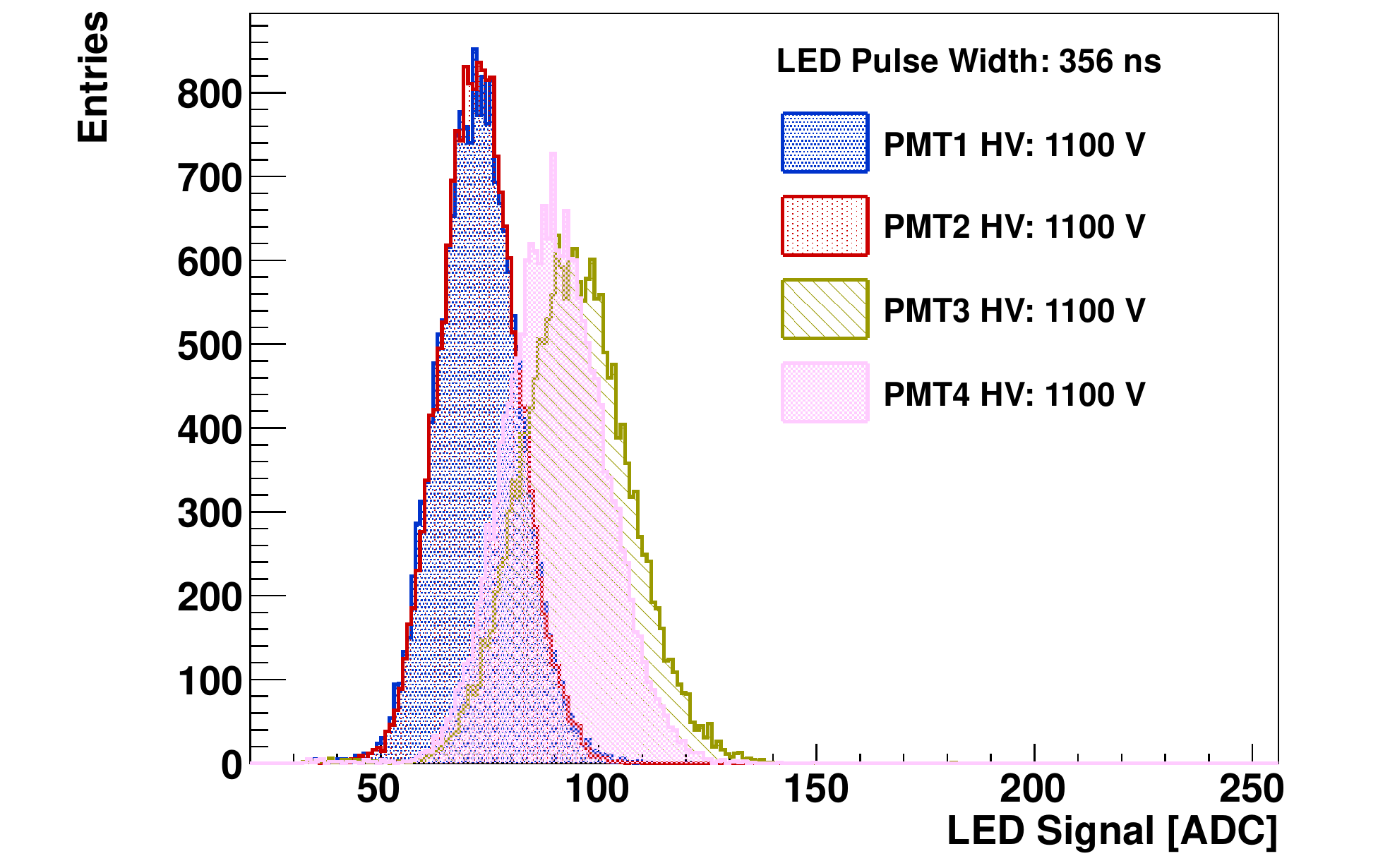}
 \hspace{-0.5cm}
 \includegraphics[height=6.5cm, width=8.5cm]{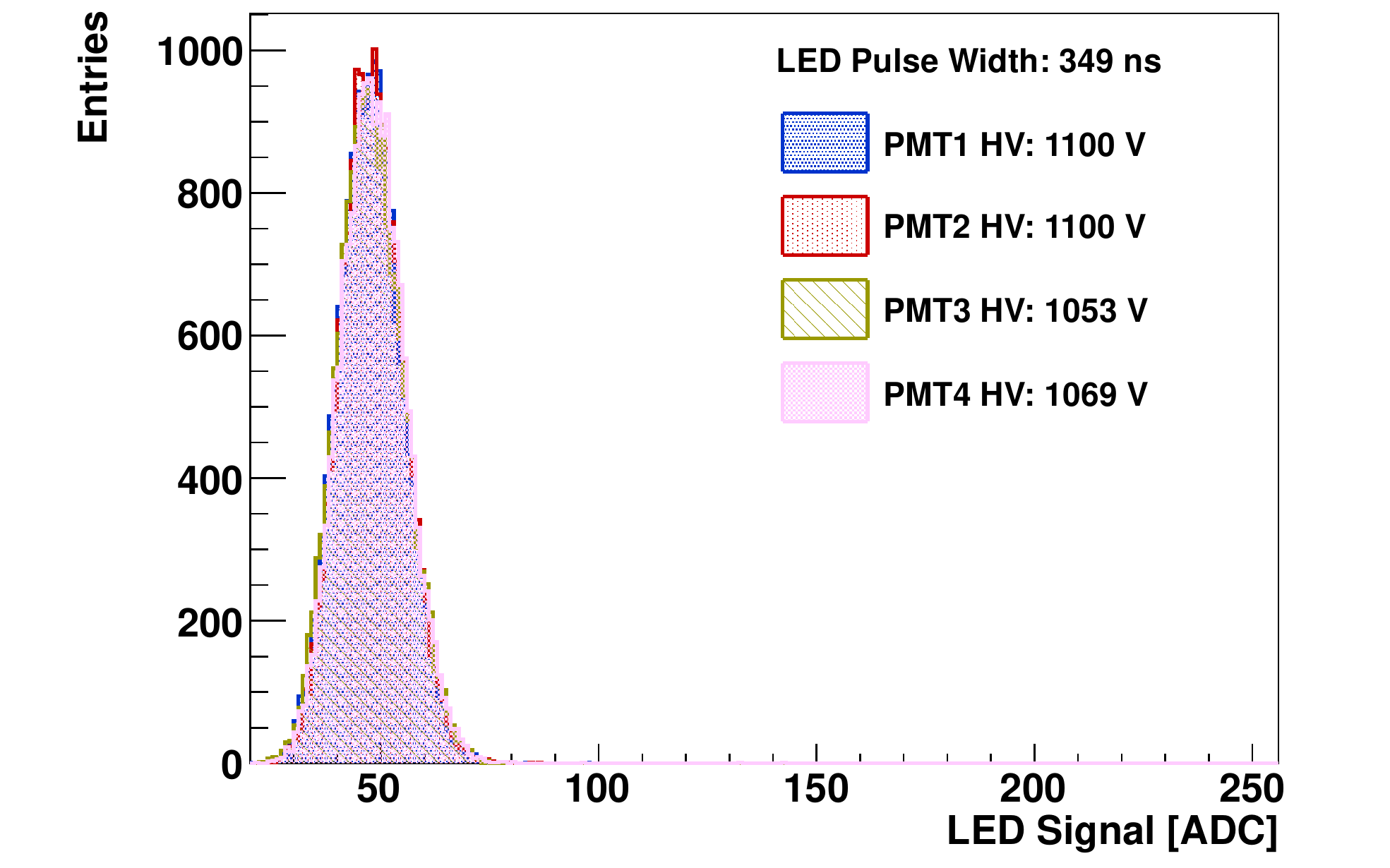}\\

 \hspace{-2.5cm}
 \includegraphics[height=6.5cm, width=8.5cm]{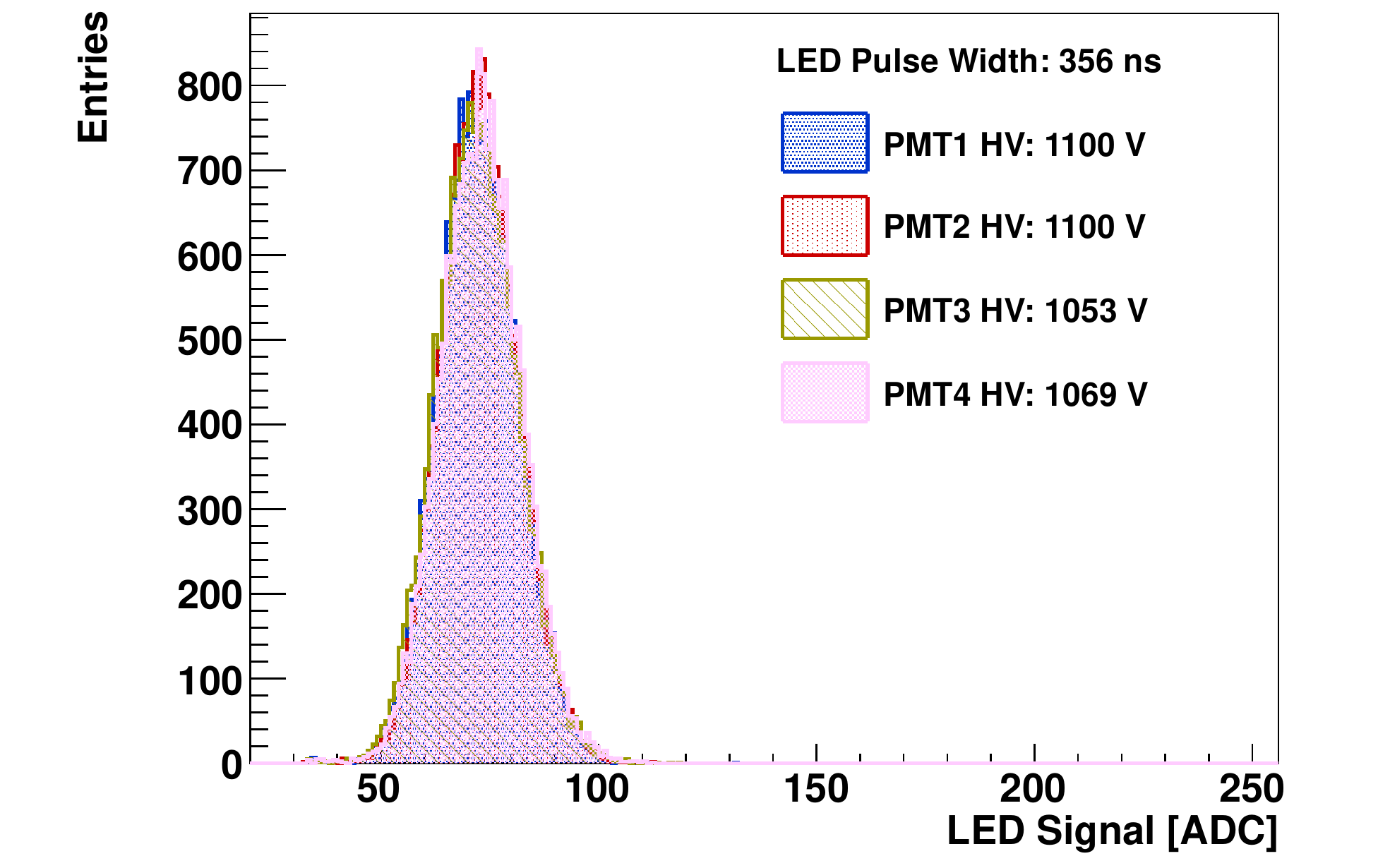}
 \hspace{-0.5cm}
 \includegraphics[height=6.5cm, width=8.5cm]{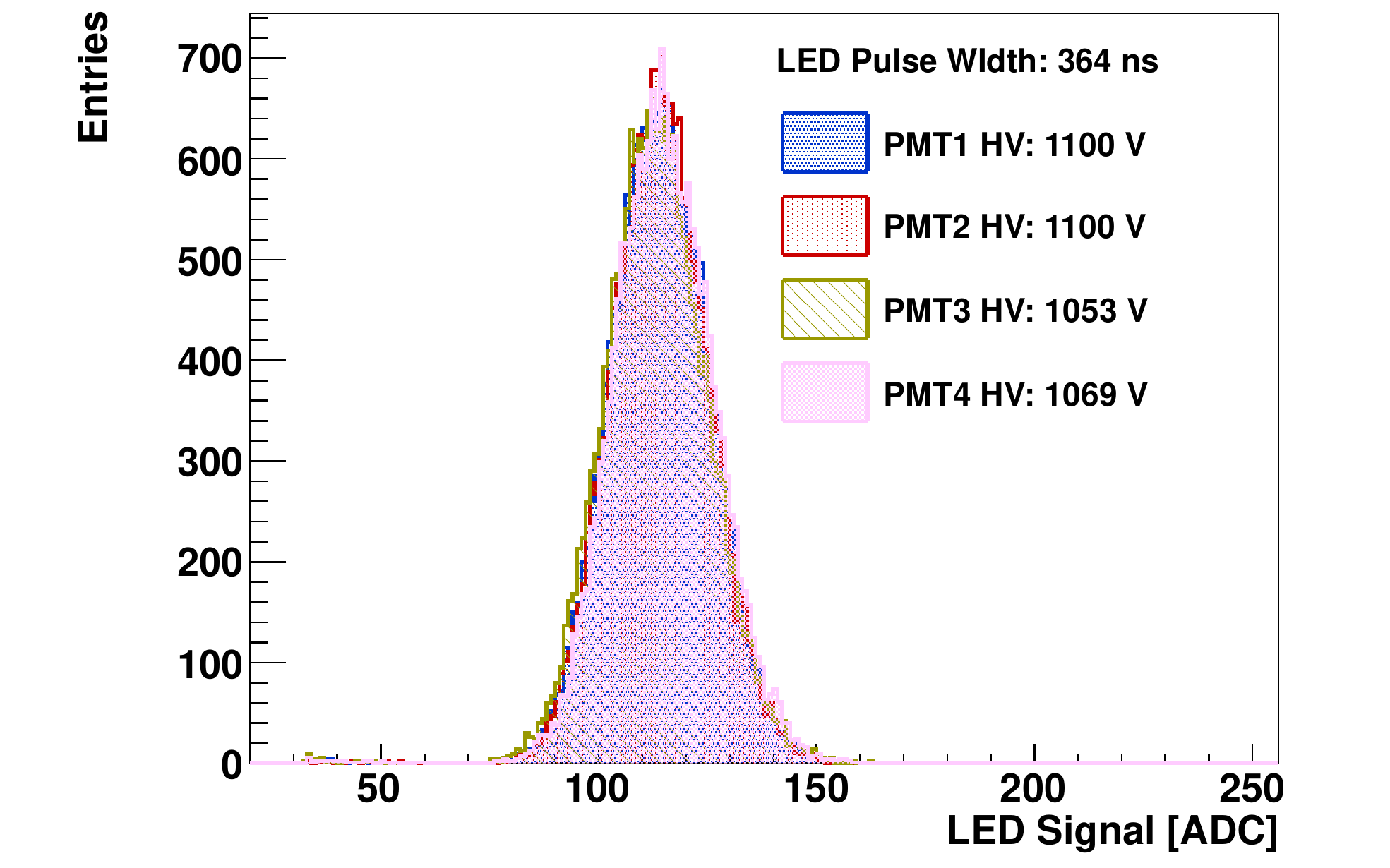}
  \vspace{-0.6cm}
  \caption{Channel Gain Matching procedure. Top Left Panel: 
           Using an LED pulse the distributions of the signal amplitude
           from the four PMTs (operated at equal high voltage value) are
           found to be displaced, although 
           the system is running with the same configuration 
           in all channels; the effect is due to the intrinsic 
           gain mismatch among channels as described in the text. 
           The remaining panels show how a gain matching can be 
           obtained properly tuning the high voltage of the PMTs,
           while remaining relatively stable upon a variation of the
           LED light intensity.}
  \label{fig:GAIN_MATCH}
\end{figure}

When properly tuning the high voltage value of the individual
PMTs, the corresponding signal distributions can be acquired 
with similar mean values and width (RMS), recovering any possible 
initial gain mismatch of the different channels of the system. 
An example of this procedure, after proper baseline correction 
performed as described above, is presented in Fig.~\ref{fig:GAIN_MATCH},
using a positive pulse with amplitude $2.96$ V.
While properly tuning the high voltage of the photomultipliers a gain 
matching among the channels can be obtained. The stability of the 
procedure against unexpected effects is clearly visible by modifying the 
width of the pulse to the LED, and observing a stable gain matching.
It should be also mentioned that this procedure to gain-match the 
readout channel should be, in principle, re-done whenever the system 
is modified (e.g. after the replacement of a photomultiplier, 
of a voltage divider, after the change of the pre-amplifier
configuration, and so on).

\begin{figure}[t!]
 \hspace{-2.5cm}
 \includegraphics[height=6.5cm, width=8.5cm]{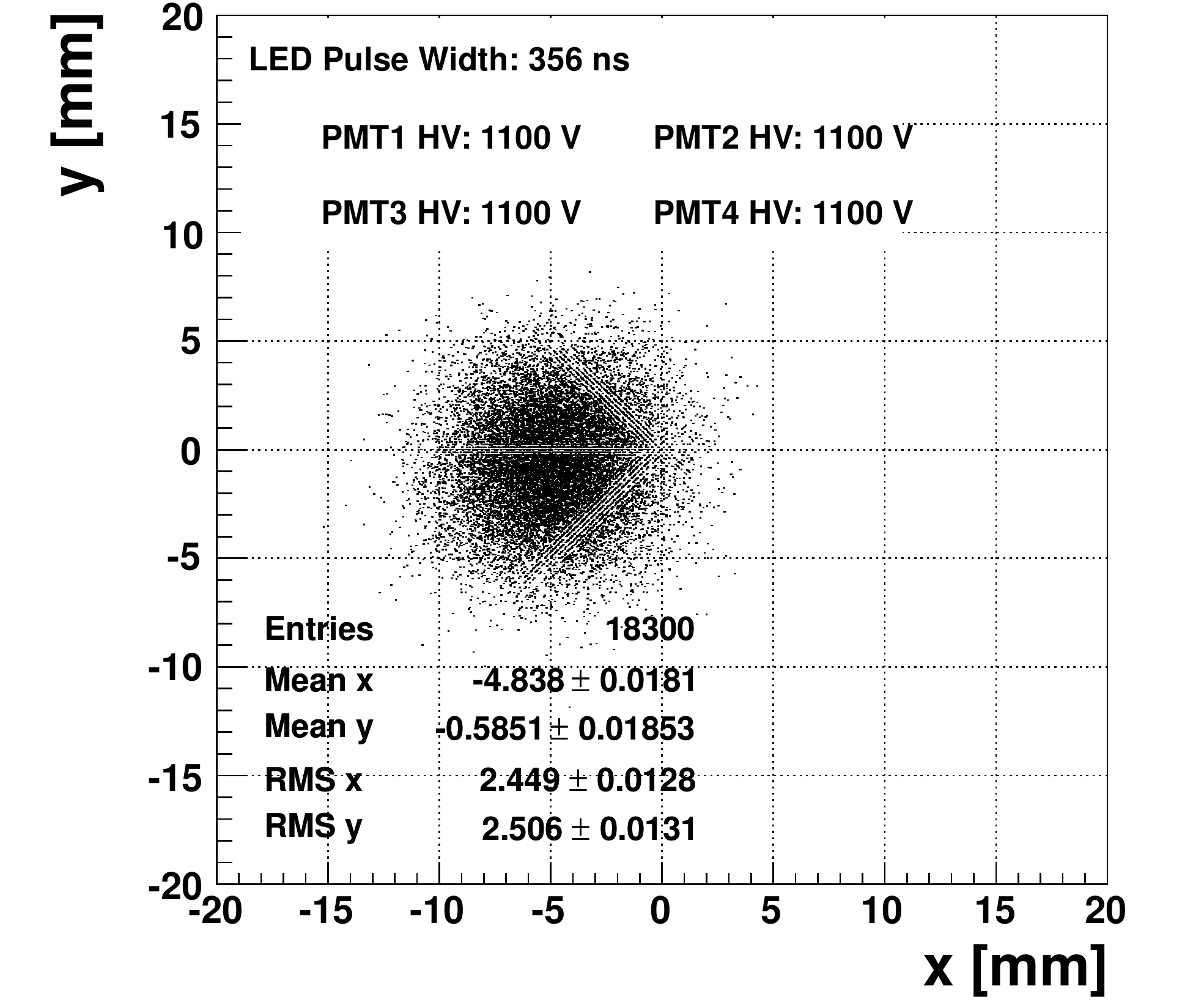}
 \hspace{-0.5cm}
 \includegraphics[height=6.5cm, width=8.5cm]{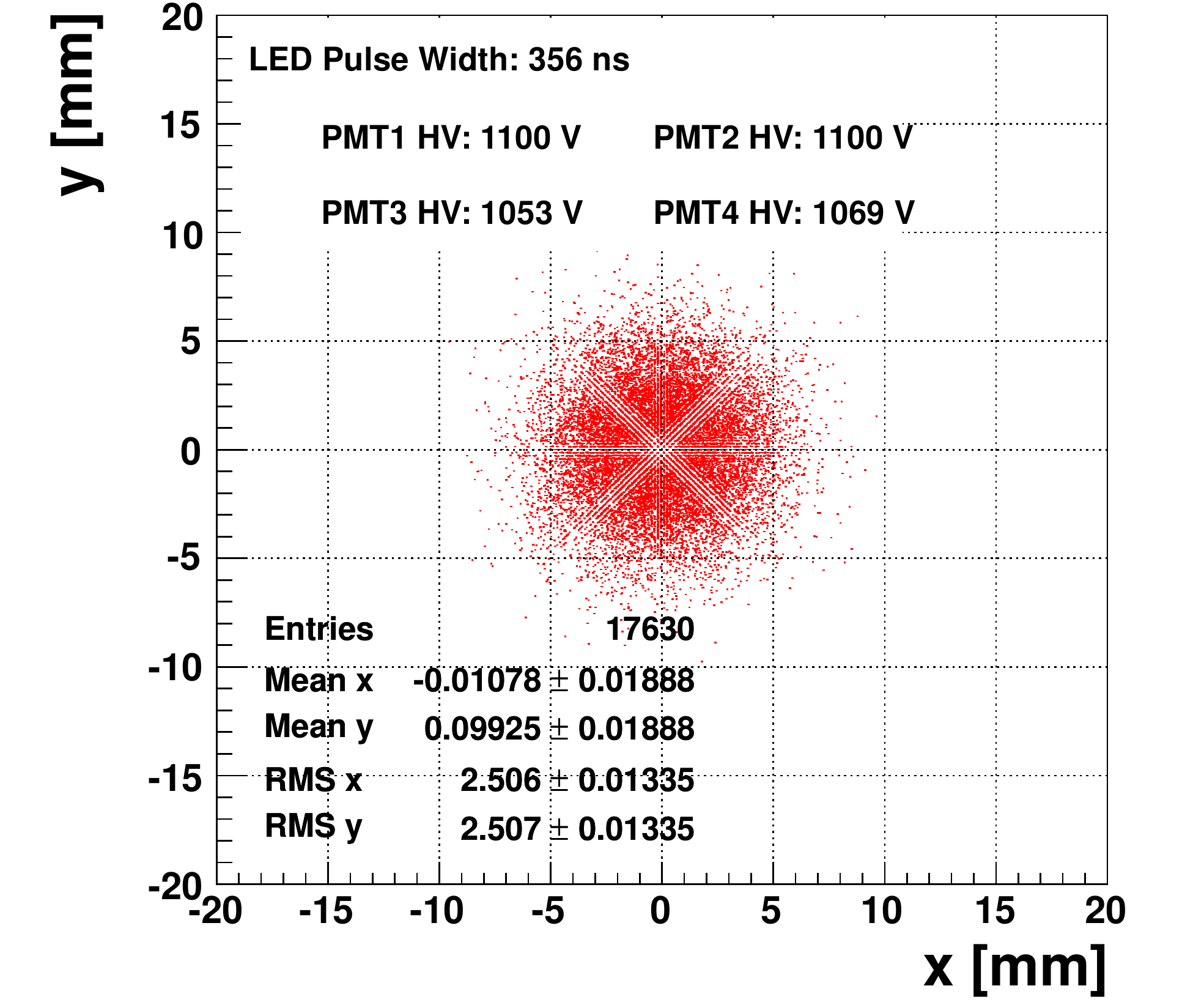}\\
  \vspace{-0.7cm}
  \caption{Effects from a gain matching procedure in the reconstruction 
           of center of gravity of LED-induced photon showers, 
           as described in the text. 
           Left Panel: 
           The high voltage value is the same for of all PMTs, regardless
           of their different intrinsic gain. Right Panel: 
           After gain matching the reconstructed showers appear to be more
           centered with respect to the PMT supporting frame.}
  \label{fig:GAIN_COG}
\end{figure}

The 2D distribution of the reconstructed $x$ and $y$ coordinates for 
the LED-generated photons is presented in Fig.~\ref{fig:GAIN_COG}
confirming (qualitatively) the goodness of the presented gain matching
procedure. 
The coordinates here presented were obtained using the "center of gravity"
method (COG) developed following this procedure: A left-right and top-down 
signal asymmetry is calculated from the response $S$ of each photomultiplier:
\begin{eqnarray}
    \eta_x & = & \frac{ S_1 + S_2 - S_3 - S_4 } { S_1 + S_2 + S_3 + S_4 } 
     \nonumber \ , \\
     & & \nonumber \\
    \eta_y & = & \frac{ S_1 + S_4 - S_2 - S_3 } { S_1 + S_2 + S_3 + S_4 }\ .
    \label{eq:etas}
\end{eqnarray}
Out of the signal asymmetries the coordinates are then calculated 
as~\cite{THESIS}
\begin{eqnarray}
    x & = & D / 2. \cdot \eta_x \nonumber \ , \\
     & & \nonumber \\
    y & = & D / 2. \cdot \eta_y \ ;
    \label{eq:xy}
\end{eqnarray}
$D \approx 80$ mm being the side width of the virtual square 
enclosing the four photomultipliers. The photomultipliers 
PMT1 to PMT4 are located anticlock-wise starting 
from the top-right corner of the holding frame.  
More sophisticated methods are available in the literature~\cite{ANTS, LPOL}.

It is interesting to stress that this LED system could be used 
(with a slight modification of the DAQ software)
to automatically monitor online any gain change 
(both long and short-term) 
providing the user a way to promptly act to correct for the
observed gain mismatch and avoiding a bias of the measurement.

The pre-amplifier was configured to provide a small and Gaussian 
signal distribution from the LED in the readout while, at the 
same time, to keep the signal distribution from neutrons well within
the ADC dynamic range when running the detector amplification 
stages at its nominal values. 
The decaying time of the output signal in the configuration 
used in our laboratory is \mbox{$C \cdot R \approx 18 \ \mu$s},
being \mbox{$C = 360$} pF and $R = 50$ kOhm.
When needed, the LED distribution can be moved at larger ADC values
by modifying the width and amplitude of the voltage pulse 
energizing the LED, as discussed in Fig.~\ref{fig:GAIN_MATCH}.

\section{Towards Real Data Taking Conditions}
%
Dedicated test-beam campaigns were done by the collaboration 
to investigate the performance of the Anger Camera prototype 
in combination with the JUDIDT readout system. The results, 
obtained after a detailed detector calibration and offline analysis, 
are shown in a dedicated report~\cite{ANGER}. Here we will 
present, for example purpose, how the Camera runs in presence 
of real neutron, using the $Cf^{252}$ neutron source. 

\begin{figure}[t!]
 \hspace{-2.5cm}
 \includegraphics[height=6.5cm, width=8.5cm]{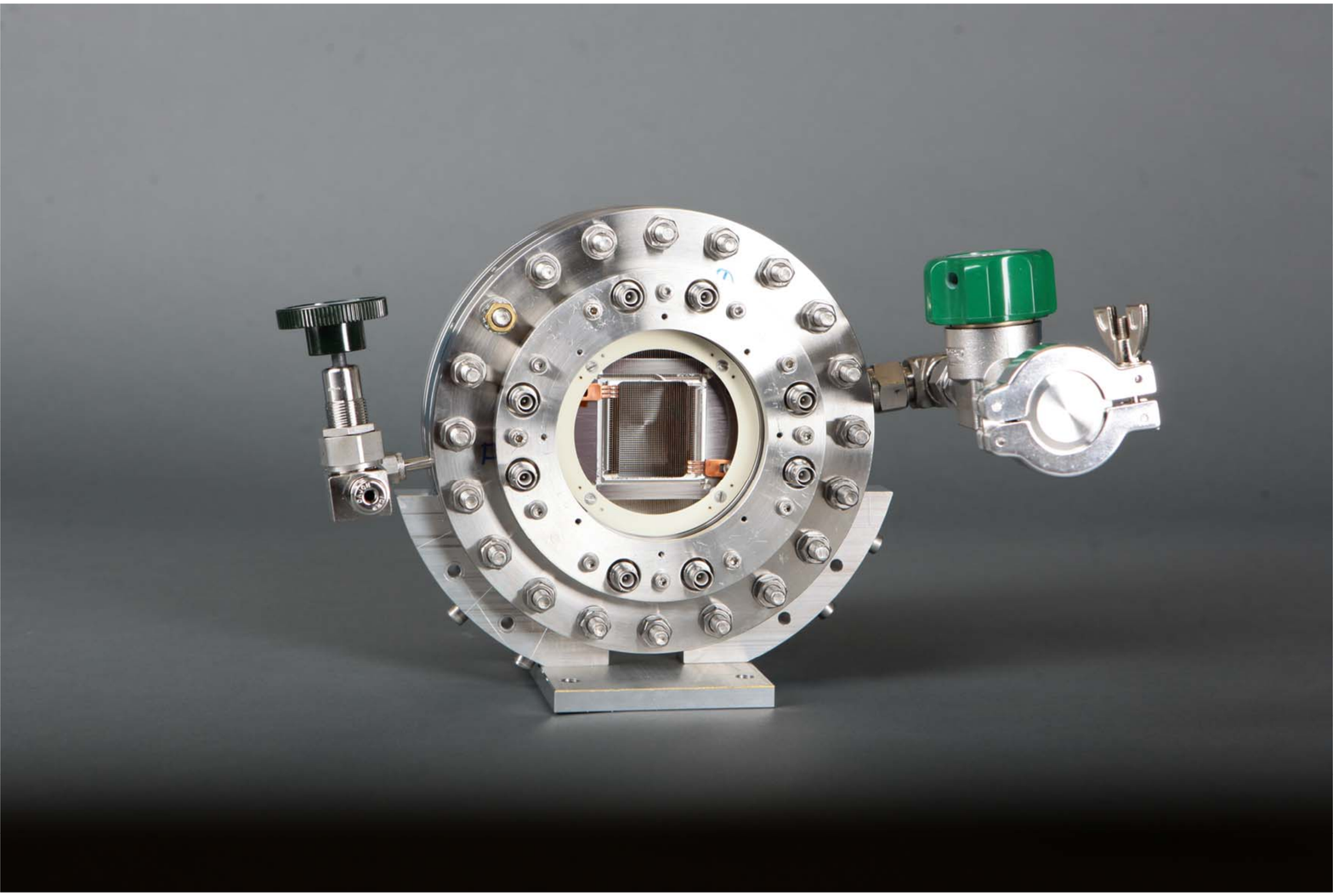}
 \hspace{0.5cm}
 \includegraphics[height=6.5cm, width=8.0cm]{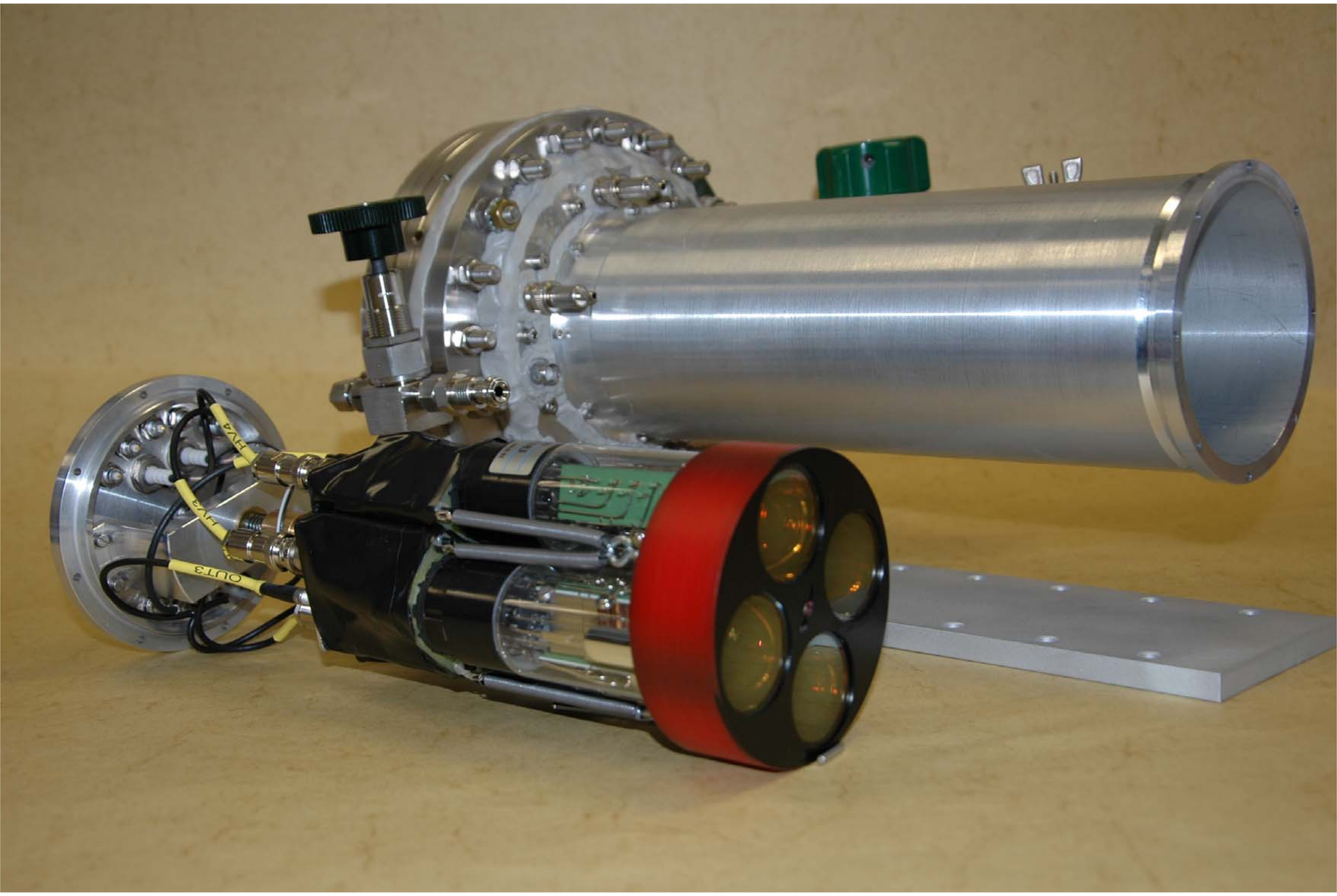}\\
  \vspace{-0.7cm}
  \caption{Left Panel: 
           Back-side view of the Anger Camera, with the micro-strip 
           amplifying stage visible in the foreground.
           Right Panel: 
           Aluminum case  hosting the photomultipliers and the 
           LED for gain-matching purposes.}
  \label{fig:ANGER_CAMERA}
\end{figure}

The camera, assembled colleagues of the 
Technische Universit\"{a}t M\"{u}nchen, 
is filled by a 
$^3He$+$CF_4$ gas mixture at the pressure of $2$ and $3$ bar, 
respectively. The back-side view is presented in the left panel 
of Fig.~\ref{fig:ANGER_CAMERA}.
Visible is the hosted micro-strip plate with an active are of 
$40$x$30$ mm$^2$, located before the exit window for the 
scintillating light. The aluminum case, shown in the right panel 
of the picture, was designed and assembled at ZEA-2. 
It is quite flexible, allowing for hosting different types of 
PMTs. 

The drawback of this flexible configuration is that the coverage 
of the amplifying micro-strip by any of the four photomultipliers can be 
different according to the orientation of the mask supporting 
the PMTs in the aluminum case, thus possibly biasing the 
measurement of the neutron impinging point. It is clear that in
case of a running experiment a fest structure will be used 
to host the PMTs, 
such to cover uniformly the active area of the micro-strip
which is responsible eventually of the largest amount of the 
scintillation process. 

\begin{figure}[b!]
 \hspace{-2.5cm}
 \includegraphics[height=6.5cm, width=8.5cm]{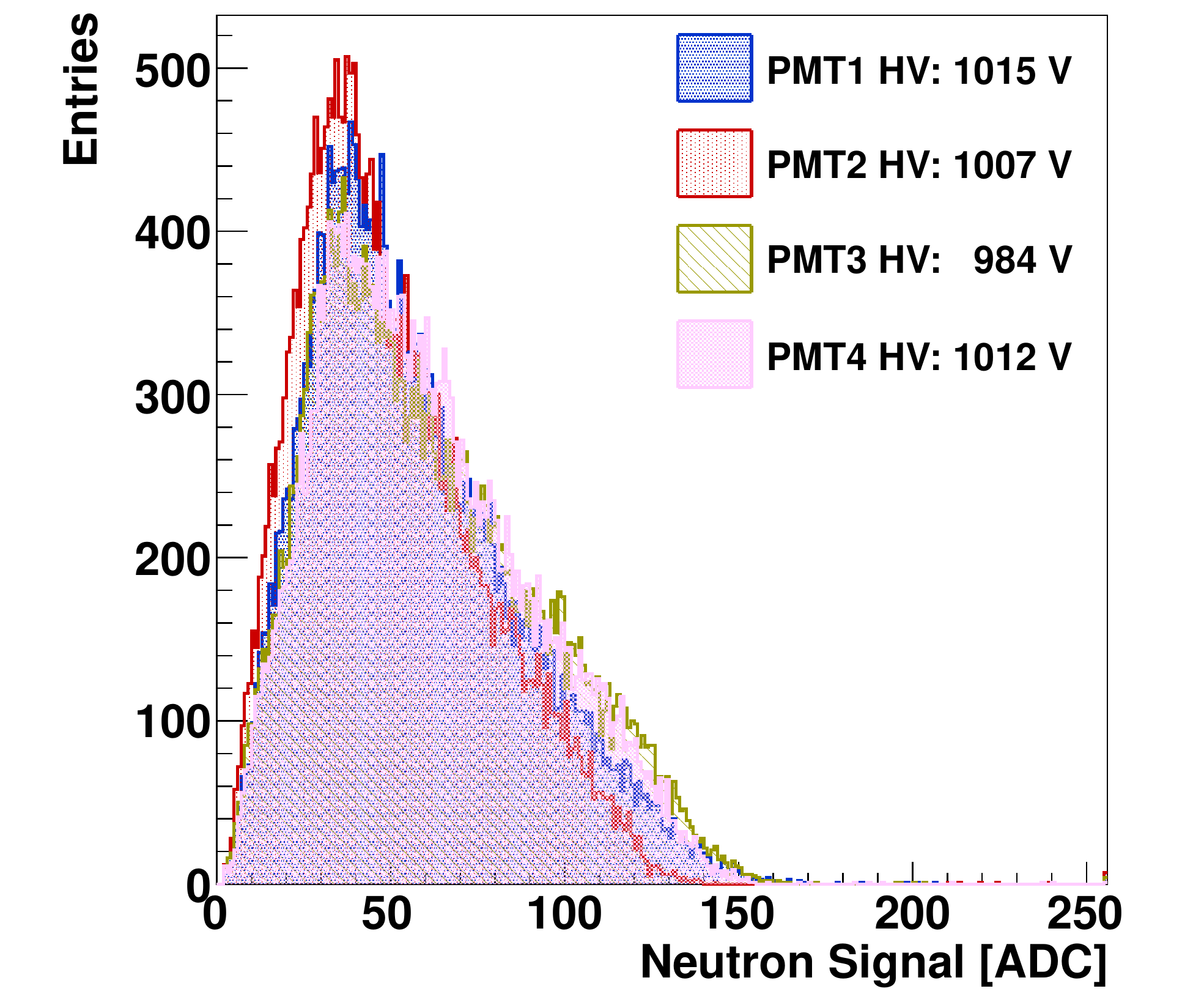}
 \hspace{0.5cm}
 \includegraphics[height=6.5cm, width=8.0cm]{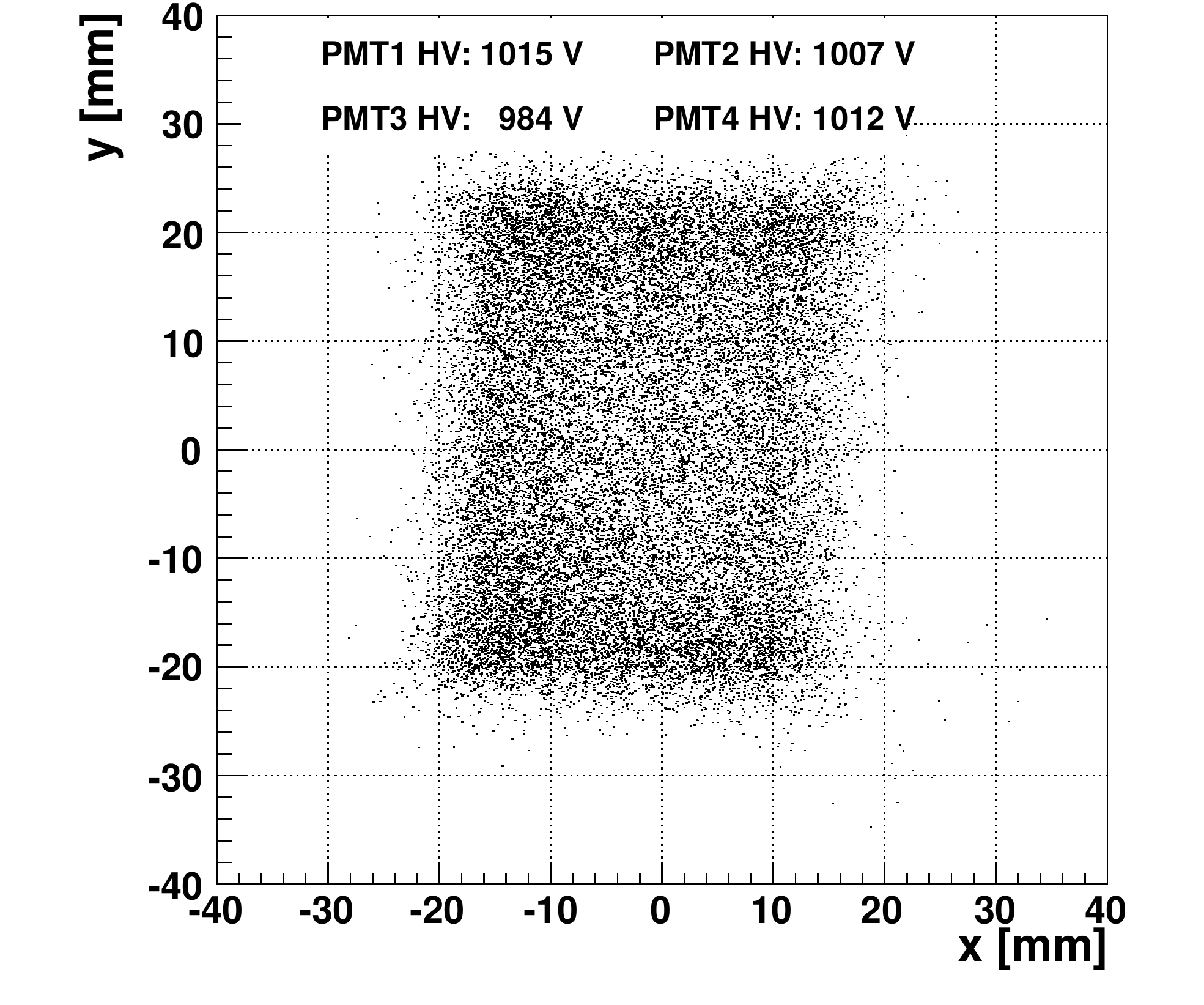}\\
  \vspace{-0.7cm}
  \caption{Left Panel: 
           Back-side view of the Anger Camera, with the micro-strip 
           amplifying stage visible in the foreground.
           Right Panel: 
           Aluminum case hosting the photomultipliers and the 
           LED for gain-matching purposes.}
  \label{fig:NEUTRON_SIGNAL}
\end{figure}

An example of the signal deposited in the chamber by neutrons 
via scintillating light is shown in the left panel of  
Fig.~\ref{fig:NEUTRON_SIGNAL}.
In the laboratory the PMT mask was oriented visually to have a 
grid of 2x2 light collectors. The detector was located at 
approximately $50$ cm from a $\phi$-symmetrically radiating neutron
source. Possibly due to the above mentioned reason,
a small residual signal asymmetry is still visible in one channel, 
as shown in the left panel of Fig.~\ref{fig:NEUTRON_SIGNAL}  
(PMT2, which is located in the bottom right corner along the neutron 
flux direction). Although the not perfect configuration 
of the PMT mask, the reconstruction of the neutron 
impinging points reproduces quite well the shape of the micro-strip
($30$x$40$ mm$^2$), shown in the right panel of the picture. 
Here the coordinate frame was used with respect to the neutron flux.
Note that also the small-degree tilt of the plate, which is due 
to mounting reasons, is reproduced in the analysis.
%

\section{Conclusions}
During the year 2012 deep systematic studies have been 
performed to characterize the readout electronics JUDIDT 
developed at Forschungszentrum J\"{u}lich, designed 
for instruments for neutron detection, in particular 
as a readout system for an  Anger Camera 
prototype filled with $^3He$-$CF_4$ gas mixture.

The more significant results are presented and described in 
this paper, showing that the overall performance of the 
electronics was found to be excellent, with an overall signal 
resolution typically around $0.3$ mV per ADC.
The intrinsic electronic noise was investigated, and was 
found to be 
within fractions of millivolt. The charge leakage between 
neighboring channels is within one ADC channel. 

Comparing the $16$ input channels of a single board, 
the pedestals showed a variable spread, which can be 
corrected for, using the flexible way of the system 
to cover a specific dynamic range 
by steering via software the baseline, individually for each 
channel, thus providing a similar dynamic range for all channels.

Several measurements, including the ADC calibration, confirmed
a linear response of the system in all the allowed dynamic 
ADC range. 

When used as readout system for the Anger Camera optimal 
spatial resolution below one mm for the neutron impact points 
was obtained, and a reasonable dead time measured up to  
$200$ kHz~\cite{ANGER}.  

All these measurements have basically fulfilled the requirements 
to operate the electronics in a neutron beam environment, and the features 
which could be fixed in the next version of the DAQ software
(as the automation of some monitoring and calibration procedures). 

At the end of 2013, the Anger Camera detector with the JUDIDT 
readout system is proposed to be used on a regular 
neutron instrument in the FRM-II experimental hall, for further 
evaluation and real scientific experiments. Having the detector 
operational on a real instrument is the ultimate test to convince 
potential users of its performance and reliability. We expect that 
several scientists will be willing to equip their instrument with a 
copy of this detector.

%
\begin{quotation}
      \vspace{0.5cm}
  \begin{center}
      {\bf Acknowledgments}
      \vspace{0.25cm}
  \end{center}
 The authors gratefully acknowledge R.\ Foester, T.\ Kollmann, 
 \mbox{C.\ Wesolek}, and the colleagues of the ZEA-2 Workshop 
 for their valuable technical contribution to the results here 
 presented.
 We are also deeply grateful to the ZEA-2 Management for its significant 
 efforts in supporting this project. 
\end{quotation}



\end{document}